\documentclass[prc,twocolumn,showpacs,preprintnumbers,amsmath,amssymb,floatfix]{revtex4}
\usepackage{graphicx}
\usepackage{float}
\usepackage{trfsigns}

\newcommand \w{{\cal O}}

\begin{document}
           \csname @twocolumnfalse\endcsname
\title{Off-shell selfenergy for 1-D Fermi liquids}
\author{Klaus Morawetz$^{1,2}$
, Vinod Ashokan$^3$, Kare Narain Pathak$^4$
}
\affiliation{$^1$M\"unster University of Applied Sciences,
Stegerwaldstrasse 39, 48565 Steinfurt, Germany}
\affiliation{$^2$International Institute of Physics- UFRN,
Campus Universit\'ario Lagoa nova,
59078-970 Natal, Brazil
}
\affiliation{$^3$Department of Physics, Dr. B. R. Ambedkar National Institute of Technology, Jalandhar 144011, India}
\affiliation{$^4$ Centre for Advanced Study in Physics, Panjab University, 160014 Chandigarh, India}

\begin{abstract}
The selfenergy in Born approximation including exchange of interacting one-dimensional systems is expressed in terms of a single integral about the potential which allows a fast and precise calculation for any potential analytically. The imaginary part of the self energy as damping of single-particle excitation's shows a rich structure of different areas limited by single-particle and collective excitation lines. The corresponding spectral function reveals a pseudogap, a splitting of excitation's into holon's and antiholon's as well as bound states.  
\end{abstract}
\maketitle

\section{Introduction}

Interacting Fermi systems in one dimension appear in many physical situations. Among them let us mention only quantum wires in carbon nanotubes \cite{Saito98,Bockrath99,Ishii03,Shiraishi03} or edge states in quantum hall liquid \cite{Milliken96,Mandal01,Chang03}, semiconducting nanowires \cite{Schafer08,Huang01}, cold atomic gases \cite{Monien98,Recati03,Moritz05}, conducting molecules \cite{Nitzan03} or even crystalline ion beams in storage rings \cite{SSH01,SSBH02}.
 
Though exact solutions are known for Luttinger \cite{L63,LP74,ES07}, Tomonaga \cite{DL74}, and Gaudin-Yang models \cite{Ram17,Pan22} of contact interaction by the Bethe ansatz \cite{EFGKK10,GBML13} the interacting Fermi system in quantum wires are still a theoretical challenge. Among these methods bosonization techniques in \cite{Lu77,Solyom79} and out of \cite{GGM10} equilibrium are employed which are based on the similar behaviour of long-distance correlations of Fermi and Bose systems \cite{H81a}. The underlying model is the continuum limit which has been discussed with the help of correlation functions \cite{Emery79}.
Already the width dependence \cite{GADMP22} of quantum wires escapes exact solutions and perturbation methods have been used to investigate analytically and numerically the ground-state properties \cite{LD11,LG16,Loos13}. The question is how relevant are results from perturbation theory for such strongly correlated one-dimensional Fermi systems. In \cite{MVBP18} it was shown that the Random Phase Approximation (RPA) becomes exact in the high-density limit for one-dimensional systems.

This peculiar feature of a perturbation series to become exact is due to the fact that in one dimension the ratio of kinetic to interaction energy is proportional to the density. Therefore the weak coupling corresponds to the high-density regime and the strong coupling regime to low densities \cite{GPS08}. Contrary to three dimensions one can therefore describe the high-density limit by a weak-coupling theory, i.e. perturbation theory. Though we cannot expect quantitative correct results by Born approximation as first-order perturbation theory of collisional damping, we might get insight into high-density correlations. Conventionally, perturbation theory is considered to fail in one dimensions due to divergences at the Fermi energy. Recently this was cured by a Pad\'e approximation and an extended quasiparticle picture does work indeed \cite{M23}. 

The selfenergy represents the fundamental quantity to study single-particle correlation and excitation effects. This is best described within the Green function technique, for an overview see \cite{V94,G04,GV08,M17b}. Green functions allow to investigate interacting models beyond exactly solvable cases and in various approximations \cite{T67,VMKA08}. The transition between Tomonaga-Luttinger and Fermi liquids has been studied\cite{Sch77,Yo01} and the resulting non-Fermi liquid behaviour has been numerically shown for Tomonaga Luttinger processes in \cite{Sch95}. For contact interaction the exact Green function has been known for a long time \cite{T67} where even the finite-size effect of the potential has been discussed. The exact impurity Green function for contact interaction was presented in \cite{Ga15}. The elastic two-particle collision in one dimension can only lead to an exchange of momenta of the two particles due to energy-momentum conservation which means that the on-shell selfenergy vanishes. 

In contrast, the off-shell selfenergy can provide an interesting insight into the physics of strongly correlated one-dimensional systems. Therefore we will provide analytical expressions for the off-shell selfenergy for electron-electron interactions in Born approximation including exchange. Due to numerous analytic reductions we present a scheme with a single integral over the potential which can be applied in a variety of situations.

The outline of the paper is as follows. Next we present the analytical result of the imaginary part of selfenergy in Born approximation including exchange. The nontrivial integration is shifted to the appendix and the particle and hole contributions to the selfenergy are discussed. Multiple ranges appear in the plot of momentum versus off-shell energy. They can be understood as originating from collective and single-particle excitation's completely nested since in one dimension the Fermi surface consists only of two points. The real part of the selfenergy as a Hilbert transform is presented in Section III where the details are again moved to the appendix. Both the imaginary and real part of the selfenergy are expressed by a single integral over any chosen potential which allows a precise and fast calculation with a wide range of applications. Taking additionally the Hartree-Fock selfenergy into account, in Section IV the spectral function is discussed up to quadratic orders in the potential or Bruckner coupling parameter. Section V summarizes and gives some conclusions. 

\section{Self energy in Born approximation}

The real part of the selfenergy is the Hilbert transform 
\begin{eqnarray} 
\sigma(k,\omega)={\rm Re}\, \sigma^R(k,\omega)=\int {d\bar \omega \over 2 \pi}{\gamma(k,\bar \omega)\over \omega-\bar \omega}
\label{Hilbert}
\end{eqnarray}
of the selfenergy spectral function or imaginary part
\begin{eqnarray}
\gamma=\sigma^> +\sigma^<=i(\sigma^R-\sigma^A)=-2 {\rm Im}\,  \sigma^R.
\label{Ga}
\end{eqnarray}
Both specifying the retarded selfenergy
\begin{eqnarray}
\sigma^R(k,\omega)=\sigma(k,\omega)-\frac i 2 \gamma(k,\omega)=\int {d\bar \omega \over 2 \pi}{\gamma(\bar \omega)\over \omega-\bar \omega+i\eta}.
\label{ret}
\end{eqnarray}
Here the electron momentum is $k$ and the off-shell energy is denoted by $\omega$. 

The retarded selfenergy determines the single-particle spectral properties of excitation's in a many-body system. The selfenergies $\sigma^\lessgtr$ describe the hole and particle contribution to the quasiparticle damping, respectively. This becomes transparent in the kinetic equation which describes the total time change of the quasiparticle distribution, $dn_k(t)/d t=I$, including a proper meanfield. From the collision integral 
\begin{eqnarray} 
I=n_k \sigma^>-(1-n_k)\sigma^<
\end{eqnarray}
one sees that $\sigma^>$ describes the contribution of damping due to particles characterized by the Fermi distribution $n_k$ in equilibrium and $\sigma^<$ the contribution to the damping due to holes $1-n_k$. The underlying Green-function formalism is quite general and it is referred in the book \cite{KB62,M17b}. Here we employ it only for elastic two-particle scattering of electrons in one dimension.

\subsection{Hole contribution to the damping}

The selfenergy in Born approximation reads
\begin{eqnarray} 
\sigma^<(k,\omega)&=&\sum\limits_{q p} 2\pi \delta (\omega\!+\!\epsilon_p\!-\!\epsilon_{p\!-\!q}\!-\!\epsilon_{k\!+\!q}) n_{p\!-\!q} n_{k\!+\!q}(1\!-\!n_p)
\nonumber\\
&& \times V_q\left [s V_q-V_{p-k-q}\right ]
\label{Born}
\end{eqnarray}
where the spin degeneracy $s$ does only apply to the direct and not to the exchange terms. In the following we understand all energies, $\omega, \gamma, \sigma$ etc, in units of Fermi energy $\epsilon_f$ and the momenta $k$ in units of Fermi momentum given by the free-particle density $n_f$  as $k_F=n_f \hbar \pi/s$. The interaction strength we express in terms of $a_B$, the Bohr-radius-equivalent \begin{eqnarray}
V_q={\hbar^2\over m a_B}v_q
\label{vq}
\end{eqnarray} 
which allows to discuss charged and neutral impurities on the same footing. The Brueckner parameter $r_s$ is the ratio of inter-particle distance $d=1/ns$ to this length  $r_s=d/a_B$. 

The implication of the occupation factors on the range of $q$-integration turns out to be quite non-trivial and are discussed in appendix~\ref{selfs}. 
We abbreviate $\Omega=\omega-k^2$ and
\begin{eqnarray}
a^2={|\Omega|\over 2}={k^2-\omega\over 2}>0
\label{a2}
\end{eqnarray}  
in the following since it is convenient when we will calculate the Hilbert transform for the real part of the selfenergy in the next Section. 

We will present the result for different momentum ranges. It can be solely given in terms of a single integral over any used potential
\begin{eqnarray}
\Phi(q)= {s^4 r_s^2\over \pi^3}\int\limits^qd \bar q{v_{\bar q}\over |\bar q|}\left (s v_q-v_{\Omega\over 2 q}\right )
\label{phi}
\end{eqnarray}
which can be calculated analytically (\ref{phia}) e.g. for contact potential $v_q=1$. For a finite-range model potential we will use
\begin{eqnarray}
v_q={1\over \sqrt{q^2+\kappa^2}}.
\label{pot}
\end{eqnarray}
Here the finite-size parameter $\kappa$ describes the width of the wire or alternatively the screening of Coulomb potential $v_q\sim 1/q$.

The result for (\ref{Born}) reads
\begin{widetext}
\begin{align}
&\underline{{\rm case }\,0<k<1:}&&&&
\nonumber\\&\quad
\sigma^<=\Phi(1-k)-\Phi\left ({a^2\over 1-k}\right )&{\rm for}
&\,0\!<\! a\!<\! 1\!-\!k & \leftrightarrow\,&  2-(k-2)^2<\omega <k^2& {\rm area:}\, 1,2,3
\nonumber\\&\quad
\sigma^<=\Phi(q_2)-\Phi(q_1)&{\rm for}&\,0< a< {1-k\over 2} & \leftrightarrow\,& {(k+1)^2\over 2}-1<\omega<k^2& {\rm area:}\, 1
\nonumber\\&\quad
\sigma^<=\Phi(1+k)-\Phi\left ({a^2\over 1+k}\right )&{\rm for}&\,0\!<\! a\!<\! 1\!+\!k& \leftrightarrow\,&  2-(k+2)^2<\omega <k^2& {\rm area:}\, 1,3,5
\nonumber\\&\quad
\sigma^<=\Phi(\bar q_2)-\Phi(\bar q_1)&{\rm for}&\,0< a< {1+k\over 2} & \leftrightarrow\,& {(k-1)^2\over 2}-1<\omega<k^2& {\rm area:}\, 1,2,3,4,5
\nonumber\\&
\underline{{\rm case }\,1<k<3:}&&&&
\nonumber\\&\quad
\sigma^<=\Phi\left ({a^2\over -1+k}\right )-\Phi(k-1)&{\rm for}&\,\sqrt{-2(1-k)}\!<\! a\!<\! \sqrt{k^2-1}& \leftrightarrow\,& 2-k^2<\omega<(k-2)^2& {\rm area:}\, 6,7
\nonumber\\&\quad
\sigma^<=\Phi(1+k)-\Phi\left ({a^2\over 1+k}\right )&{\rm for}&\,\sqrt{k^2-1}\!<\! a\!<\! {1\!+\!k}& \leftrightarrow\,& 2-(k+2)^2<\omega<2-k^2& {\rm area:}\, 4,5
\nonumber\\&\quad
\sigma^<=\Phi(\bar q_2)-\Phi(\bar q_1)&{\rm for}&\,\sqrt{-2(1-k)}< a< {1+k\over 2} & \leftrightarrow\,& {(k-1)^2\over 2}-1<\omega<(k-2)^2& {\rm area:}\, 5,6
\nonumber\\&
\underline{{\rm case }\,3<k:}&&&&
\nonumber\\&\quad
\sigma^<=\Phi\left ({a^2\over -1+k}\right )-\Phi(-1+k)&{\rm for}&\,1-k\!<\! a\!<\! \sqrt{k^2-1}& \leftrightarrow\,& 2-k^2<\omega<2-(k-2)^2& {\rm area:}\, 7
\nonumber\\&\quad
\sigma^<=\Phi(1+k)-\Phi\left ({a^2\over 1+k}\right )&{\rm for}&\,\sqrt{k^2-1}\!<\! a\!<\! 1+k& \leftrightarrow\,& 2-(k+2)^2<\omega<2-k^2& {\rm area:}\, 4
\label{forma}
\end{align}
\end{widetext}
with (\ref{q12})
\begin{eqnarray}
q_{1/2}={1-k\over 2}\pm \sqrt{{(1-k)^2\over 4}-a^2}.
\label{q12a}
\end{eqnarray}
and $\bar q_{1/2}$ are (\ref{q12a}) with $k\to -k$.

\begin{figure}[h]
\includegraphics[width=9cm]{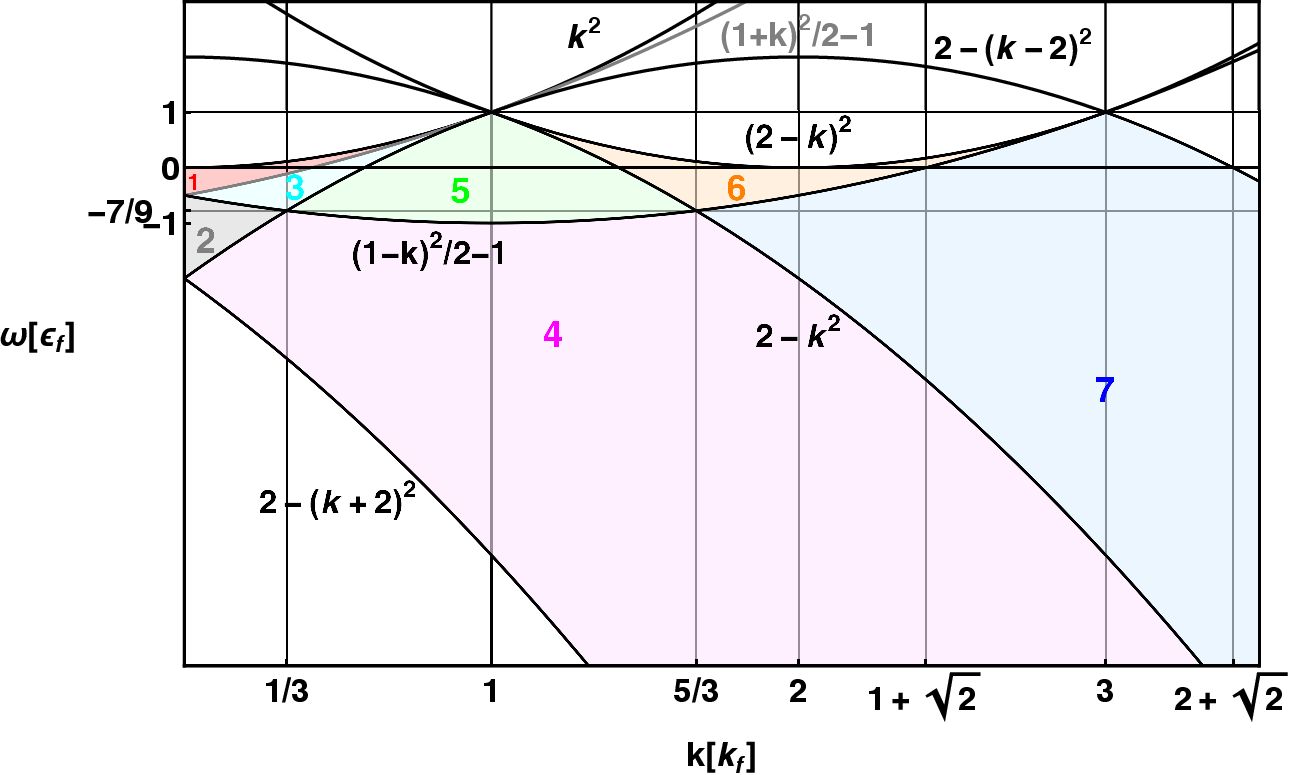}
\caption{
The 7 different areas for the selfenergy $\sigma^<$ according to (\ref{forma}). The crossing of characteristic lines at the special points $k=1,3$ are visible.
\label{wk_smaller}
}
\end{figure}

One sees that the analytical expression distinguishes three areas with respect to the momentum $k$. For momentum smaller than the Fermi momentum $k<1$ we can imagine that a second particle is situated inside the Fermi sphere of another electron. In the range $1<k<3$ the electron's Fermi sphere $k+1$ can be considered inside the Fermi interval $k\pm 1$ of a second particle, i.e. ranging from 1 to 3.

Discussing the forms with respect to momentum and frequency, these momentum areas are more involved. 
Seven different areas  in frequency $\omega$ appear and are plotted in figure~\ref{wk_smaller}. The border curves between these areas have a physical meaning. The part $\sigma^<$ describes the damping due to holes and lays below the on-shell $\omega=k^2$. Since the one-dimensional system is maximally nested we do have all borders also with $k\to k\pm2$. i.e. also below the shell $(k-2)^2$. 

The borders $(1\pm k)^2/2-1$ can be understood as arising from the collective behaviour. Since the Fermi surface shrinks at T = 0 to two points the
single-particle excitation turns into collective one \cite{SDC19}. These collective excitation's are described by the polarization function which determines the dielectric function. The lowest-order polarization in RPA reads 
\begin{eqnarray}
\Pi_0(q,\omega)={m s\over \pi p_f \hbar}\int{dp} {n_{p+\frac q 2}-n_{p-\frac q 2}\over (p+\frac q 2)^2-(p-\frac q 2)^2-\omega-i0}
\end{eqnarray}
which becomes for zero temperature \cite{MVBP18,VBMP19}
\begin{eqnarray}
{\rm Im} \Pi_0&=&-{m s\over 2 \hbar p_f q}\Theta(\omega-|\omega_-|)\Theta(|\omega_+|-\omega),\nonumber\\
{\rm Re} \Pi_0&=&{m s\over 2 \pi \hbar p_f q} \ln{\left |{\omega^2-\omega_-^2\over \omega^2-\omega_+^2}\right |}
\label{p0t}
\end{eqnarray}
with $\omega_\pm={q}(q \pm 2)$ in units of Fermi energy. This gives the limiting line where collective excitation's occur. Subtracting the Fermi energy and considering the reduced mass due to two-particle scattering translates into $(1\pm q)^2/2-1$ lines of figure~\ref{wk_smaller}. These borders indicate the divergence of polarization known as Kohn anomaly and which are the reason for Peierl's instability.

\begin{figure}[h]
\includegraphics[width=6cm]{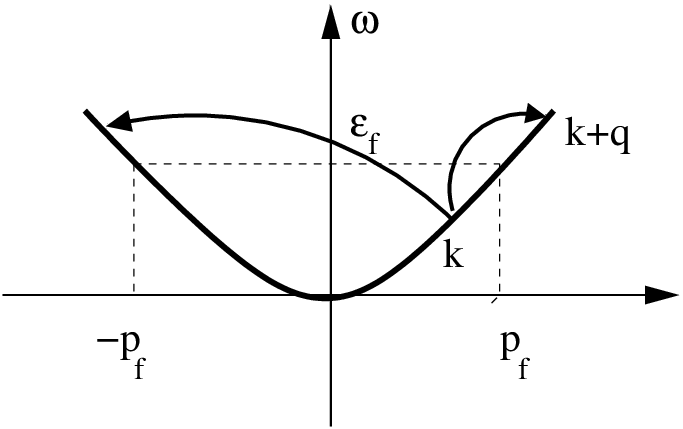}
\caption{The scheme of possible single-particle excitation due to scattering out of Fermi sea.
\label{exit}}
\end{figure}

The second class of lines are $2-k^2$, or nested as $2-(k\pm 2)^2$, arising from single particle excitation \cite{Pan22} due to off-shell scattering. For explanation we consider a simple model in figure~\ref{exit} similar to \cite{G04}. The excitation of a particle with momentum $k$ out of the Fermi sea $\omega_k(q)=\epsilon_{k+q}-\epsilon_k$ due to scattering with momentum $q$ arises for possible particle momenta $k\in (k_f-q,k_f)$ for $k<2 k_f$ and $k\in (-k_f,k_f)$ for $k>2 k_f$. Averaging this excitation $w_k(q)$ about the possible interval $k$ and choosing as fluctuation the maximum and minimum possible excitation in this interval we obtain in units of Fermi energy 
\begin{eqnarray}
\omega(q)=\left \{ \begin{array}{cc}2q\pm 2 q^2;&q<2\cr q^2\pm 4 q;&q>2\end{array}\right .\end{eqnarray}   
which second case yields after subtracting from the two Fermion threshold the curves $2-(k\pm2)^2$ in figure~\ref{wk_smaller}. 

\begin{figure}[h]
\includegraphics[width=4.2cm]{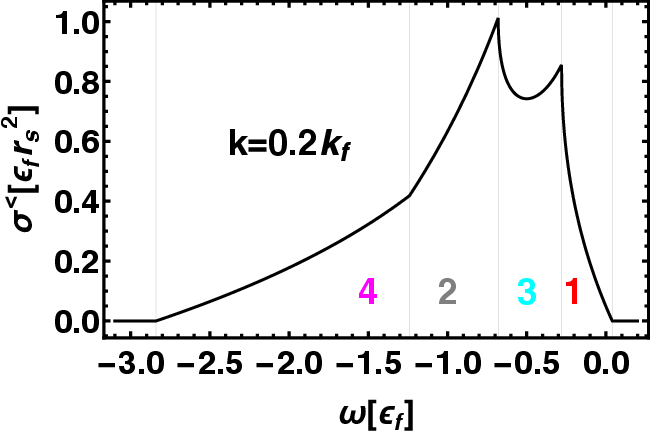}
\includegraphics[width=4.2cm]{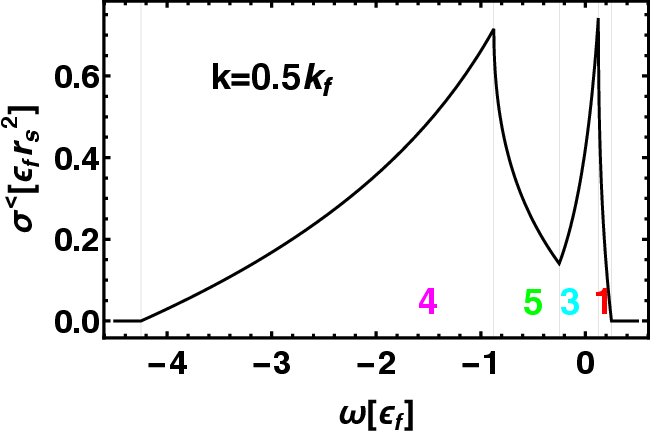}
\\
\includegraphics[width=4.2cm]{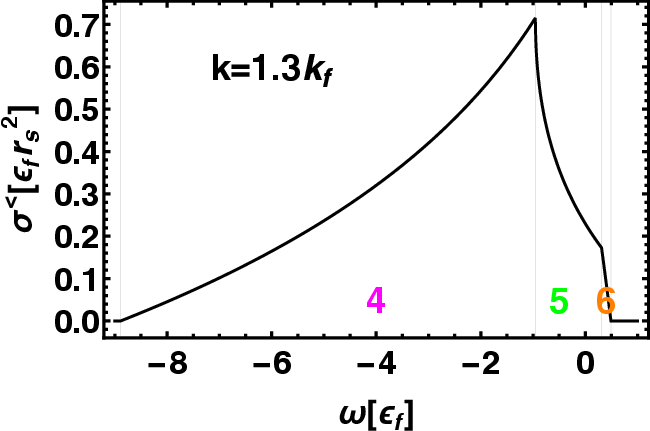}
\includegraphics[width=4.2cm]{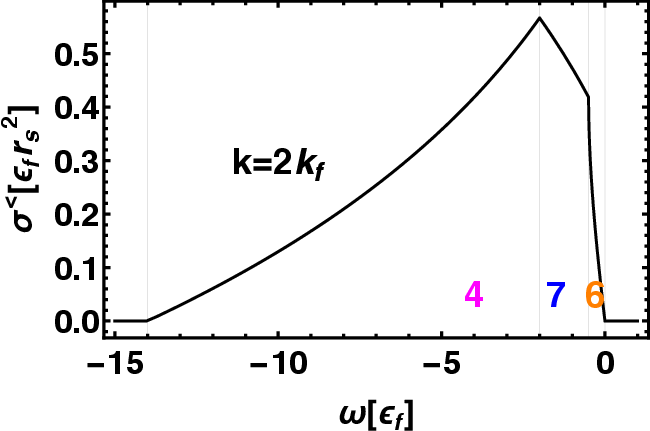}
\\
\includegraphics[width=4.2cm]{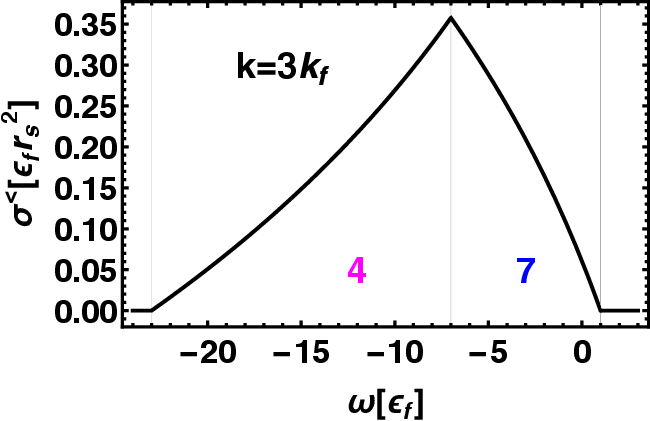}
\caption{The selfenergy $\sigma^<$ of contact interaction for different momentum cuts and regions numbers of figure~\ref{wk_smaller}.
\label{ss_k}}
\end{figure}

For exploratory reasons we use contact interaction $v_q=1$ since the limits of the ranges are independent of the interaction. Only the quantitative value of the selfenergy insides these ranges will change with the potential which we discuss later. We plot in figures~\ref{ss_k} the selfenergy $\sigma^<$ for different momenta cuts according to figure~\ref{wk_smaller} covering the crossing of various areas by frequency.
One sees a continuous but non-differentiable behaviour.

\subsection{Particle contribution to the damping}

The particle contribution $\sigma^>$ to the selfenergy is obtained from (\ref{Born}) by interchanging the distribution or occupation factors $n\leftrightarrow 1-n$.  Therefore the second part of the damping (\ref{Ga}) now due to particles reads
\begin{eqnarray}
\sigma^>(k,\omega)&=&\sum\limits_{q p} V_q\left [s V_q-V_{p-k-q}\right ] 2\pi \delta (\omega\!+\!\epsilon_p\!-\!\epsilon_{p\!-\!q}\!-\!\epsilon_{k\!+\!q}) 
\nonumber\\
&& \times (1\!-\!n_{p\!-\!q})(1\!-\! n_{k\!+\!q})n_p
\label{Bornl}
\end{eqnarray}
which integration is presented in appendix~\ref{selfg}.
We have to distinguish two cases.

\begin{figure}[h]
\includegraphics[width=9cm]{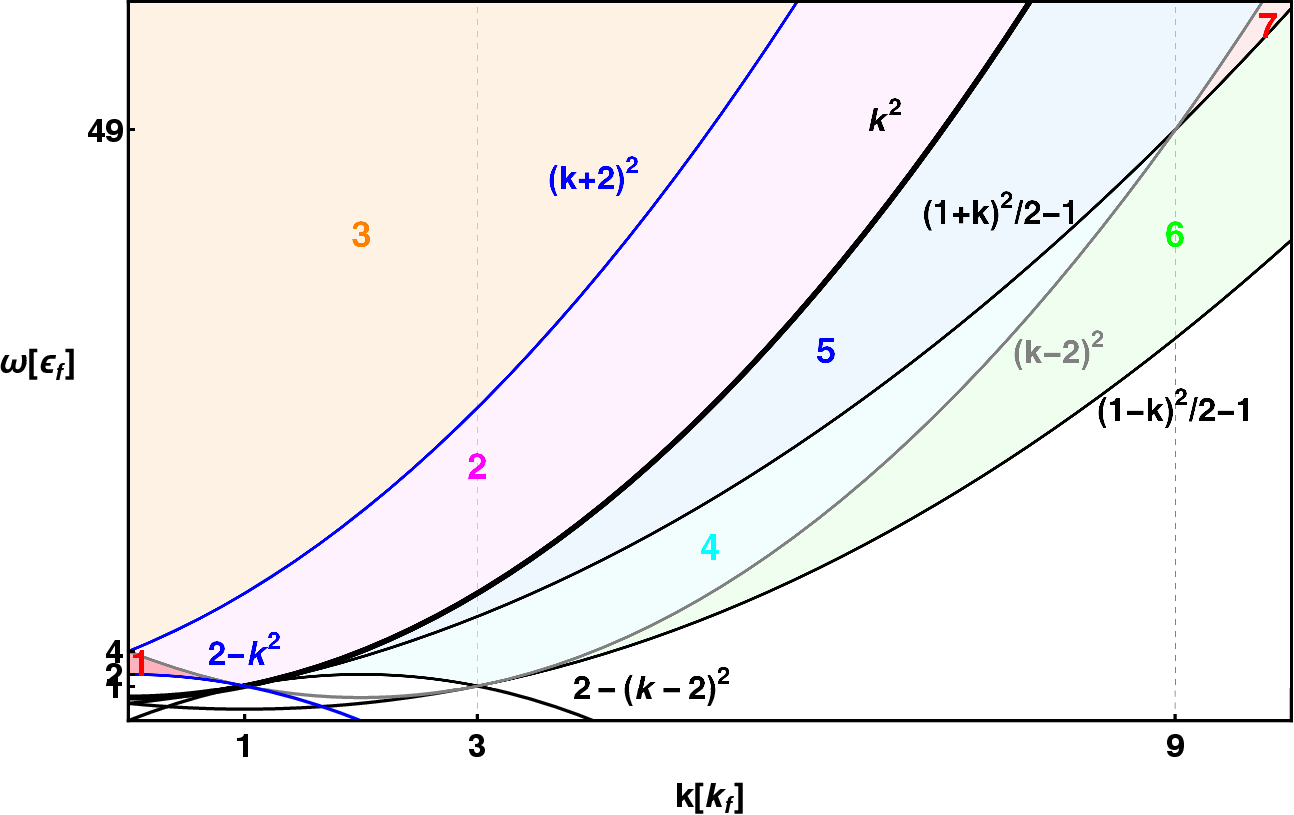}
\caption{The 7 different areas for the selfenergy $\sigma^>$ according to (\ref{formb}) and (\ref{formc}).
\label{wk_larger}}
\end{figure}

In the first case, $\Omega=\omega-k^2<0$, we obtain
\begin{widetext}
\begin{align}
&1<k<3&&&&&
\nonumber\\&\quad
\sigma^>=\Phi(-1+k)-\Phi\left ({a^2\over -1+k}\right )&{\rm for}&\,0\!<\! a\!<\! k-1& \leftrightarrow\,& 2-(k-2)^2<\omega<k^2& {\rm area:}\, 4,5
\nonumber\\&\quad
\sigma^>=\Phi(\bar q_1)-\Phi(\bar q_2)&{\rm for}&\,0\!<\! a\!<\! {(k\!-\!1)\over 2}& \leftrightarrow\,& {(k+1)^2\over 2}-1<\omega<k^2& {\rm area:}\, 5,7
\nonumber\\&
3<k&&&&&
\nonumber\\&\quad
\sigma^>=\Phi(\bar q_1)-\Phi(\bar q_2)&{\rm for}&\,0\!<\! a\!<\! {(k\!-\!1)\over 2}& \leftrightarrow\,& {(k+1)^2\over 2}-1<\omega<k^2& {\rm area:}\, 4,5,7
\nonumber\\&\quad
\sigma^>=\Phi(-1+k)-\Phi({a^2\over -1+k})&{\rm for}&\,0\!<\! a\!<\! \sqrt{2(k-1)}& \leftrightarrow\,& (k-2)^2<\omega<k^2& {\rm area:}\, 4,5
\nonumber\\&\quad
\sigma^>=\Phi(\bar q_4)-\Phi(\bar q_3)&{\rm for}&\,\sqrt{2(k-1)}\!<\! a\!<\! {1+k\over 2}& \leftrightarrow\,& {(k-1)^2\over 2}-1<\omega<(k-2)^2& {\rm area:}\, 6,7
\label{formb}
\end{align}
\end{widetext}
with (\ref{phi}), 
\begin{eqnarray}
q_{3/4}={-1-k\over 2}\pm \sqrt{{(1+k)^2\over 4}-a^2},
\label{q34a}
\end{eqnarray}
and $\bar q_{3/4}$ are (\ref{q34a}) with $k\to -k$. This case contributes to the ranges $4-7$ plotted in figure~\ref{wk_larger}.

In the second case $\Omega=\omega-k^2>0$ we obtain different areas and arguments
\begin{widetext}
\begin{align}
&0<k<1&&&&
\nonumber\\&\quad
\sigma^>=\Phi\left ({a^2\over 1+k}\right )-\Phi(k+1)&{\rm for}&\,\sqrt{1-k^2}\!<\! a\!<\!\sqrt{2(1+k)}& \leftrightarrow\,& 2-k^2<\omega<(k+2)^2& {\rm area:}\, 4
\nonumber\\&\quad
\sigma^>=\Phi\left ({a^2\over 1-k}\right )-\Phi(1-k)&{\rm for}&\,\sqrt{1-k^2}\!<\! a\!<\!\sqrt{2(1-k)}& \leftrightarrow\,& 2-k^2<\omega<(k-2)^2& {\rm area:}\, 1
\nonumber\\&\quad
\sigma^>=\Phi\left (\bar q_7\right)-\Phi(q_8)&{\rm for}&\,\sqrt{2(1-k)}\!<\! a\!<\!\infty& \leftrightarrow\,& (k-2)^2<\omega<\infty& {\rm area:}\, 2,3,4
\nonumber\\&\quad
\sigma^>=\Phi\left (q_7\right)-\Phi(\bar q_8)&{\rm for}&\,\sqrt{2(1+k)}\!<\! a\!<\!\infty& \leftrightarrow\,& (k+2)^2<\omega<\infty& {\rm area:}\, 3
\nonumber\\&
1<k&&&&
\nonumber\\&\quad
\sigma^>=\Phi\left ({a^2\over 1+k}\right )-\Phi(k+1)&{\rm for}&\,0<a<\sqrt{2(1\!+\!k)}& \leftrightarrow\,& k^2<\omega<(k+2)^2& {\rm area:}\, 4
\nonumber\\&\quad
\sigma^>=\Phi\left (\bar q_7\right)-\Phi(q_8)&{\rm for}&\,0<a<\infty& \leftrightarrow\,& k^2<\omega<\infty& {\rm area:}\, 2,3,4
\nonumber\\&\quad
\sigma^>=\Phi\left (q_7\right)-\Phi(\bar q_8)&{\rm for}&\,\sqrt{2(1+k)}\!<\! a\!<\!\infty& \leftrightarrow\,& {(k+1)^2\over 2}-1<\omega<k^2& {\rm area:}\, 3
\label{formc}
\end{align}
\end{widetext}
with (\ref{phi}), $a^2={|\Omega|/2}={(\omega-k^2)/2}$,
\begin{eqnarray}
q_{7/8}={\pm 1-k\over 2}+ \sqrt{{(1\mp k)^2\over 4}+a^2},
\label{q78a}
\end{eqnarray}
and $\bar q_{7/8}$ are (\ref{q78a}) with $k\to -k$.
This case contributes to the ranges $1-4$ plotted in figure~\ref{wk_larger} which summarizes the different areas of (\ref{formb}) and (\ref{formc}). The occurring border lines are the same as in figure~\ref{wk_smaller} explained there.

\begin{figure}[h]
\includegraphics[width=4.2cm]{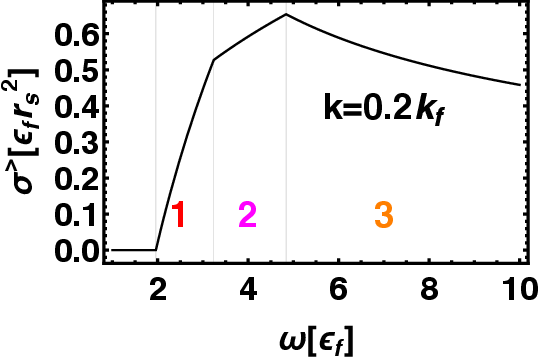}
\includegraphics[width=4.2cm]{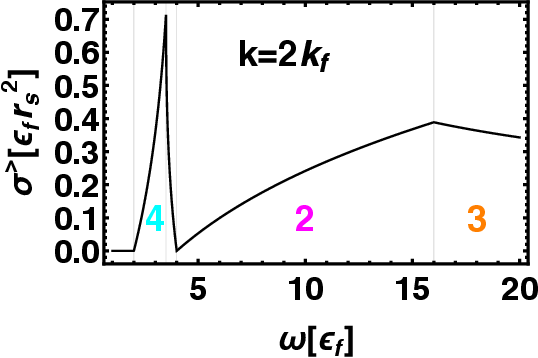}
\\
\includegraphics[width=4.2cm]{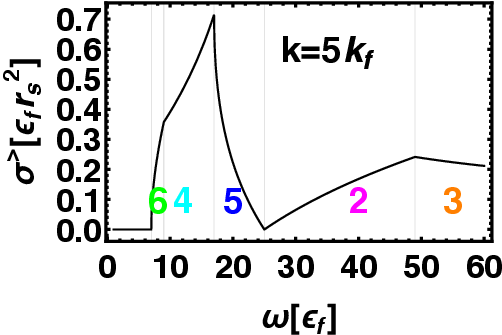}
\includegraphics[width=4.2cm]{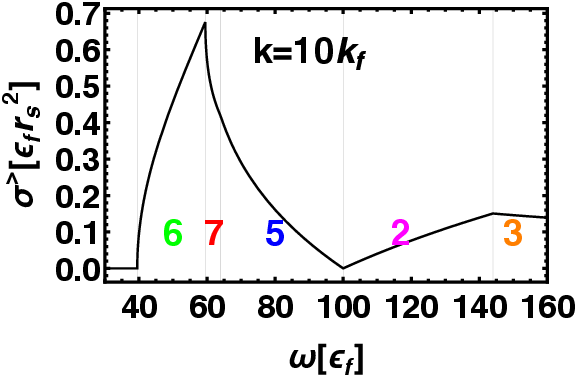}
\caption{The selfenergy $\sigma^>$ of contact interaction for different momentum cuts of figure~\ref{wk_larger}.
\label{sl_k}}
\end{figure}

In figures~\ref{sl_k} the selfenergy $\sigma^>$ is presented for different momentum cuts covering the crossing of various areas by frequency in figure~\ref{wk_larger}.
One observes again a continuous but non-differentiable behaviour.

The figures~\ref{wk_larger} and \ref{wk_smaller} together give the complete ranges of the imaginary part of the selfenergy. We like to point out that the Eq.s (\ref{forma}), (\ref{formb}) and (\ref{formc}) allow to calculate analytically this imaginary part for any interaction with the help of a single integral (\ref{phi}) which provides a fast and precise calculation.

\begin{figure}[h]
\includegraphics[width=8cm]{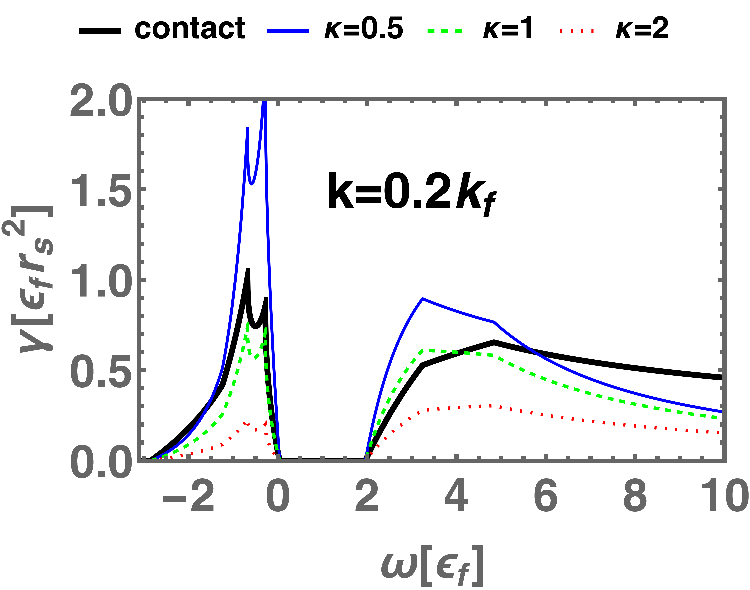}
\caption{The imaginary part of selfenergy for contact interaction and three values of finite size potential (\ref{pot}).
\label{finitei}
}
\end{figure}

The finite-size potential (\ref{pot}) with (\ref{phia}) can be used as well
and is presented in figure~\ref{finitei}. One sees exactly the same borderlines of areas as discussed above but different quantitative values dependent on the width parameter $\kappa$. For smaller $\kappa$ we approach the Coulomb potential and one sees that the peaks become enhanced. 

\section{Real part of selfenergy}

The Hilbert transform (\ref{Hilbert}) we will perform in the appendix~\ref{reals} and obtain with the help of a single integral about arbitrary potentials
\begin{align}
\phi(x)={s^5 r_s^2\over 2 \pi^4}\int\limits^x
{dx\over {\omega-k^2\over 2}+x}
\left (s v_q-v_{x\over  q}\right )
\label{psi1}
\end{align}
with the abbreviations
\begin{eqnarray}
\Theta_{ij}&=&\Psi^0_{ij}-\Psi_{ij}
\nonumber\\
\Psi_{ij}&=& \phi(i k q+j q -q^2),
\quad
\Psi^0_{ij}=\phi(i k q+j q ) 
\end{eqnarray}
finally 
\begin{align}
\sigma &=\int\limits_{0}^{2}\!\! dq{v_q\over q}\left (\Theta_{+-}\!+\!\Theta_{--}\right )\nonumber\\
&+\int\limits_{2}^{\infty}\!\! dq{v_q\over q}\left (\Psi_{++}\!+\!\Psi_{-+}-\Psi_{+-}\!-\!\Psi_{--}\right )
\nonumber\\
&
\!+\!\left \{\!
\begin{array}{cc}
\!\int\limits_{k-1}^{k+1} \!dq{v_q\over q}\, \left (\Theta_{++}\!-\!\Theta_{+-}\right )& k\!>\!1
\cr
\int\limits_{0}^{k+1}\!\! dq{v_q\over q}\left (\Theta_{++}\!-\!\Theta_{+-}\right)
\!+\!
\!\int\limits_{0}^{1-k}\!\! dq{v_q\over q}\left (\Theta_{-+}\!-\!\Theta_{--}\right )&k\!<\!1
\end{array}
\right ..
\label{real}
\end{align}

\begin{figure}[h]
\includegraphics[width=8cm]{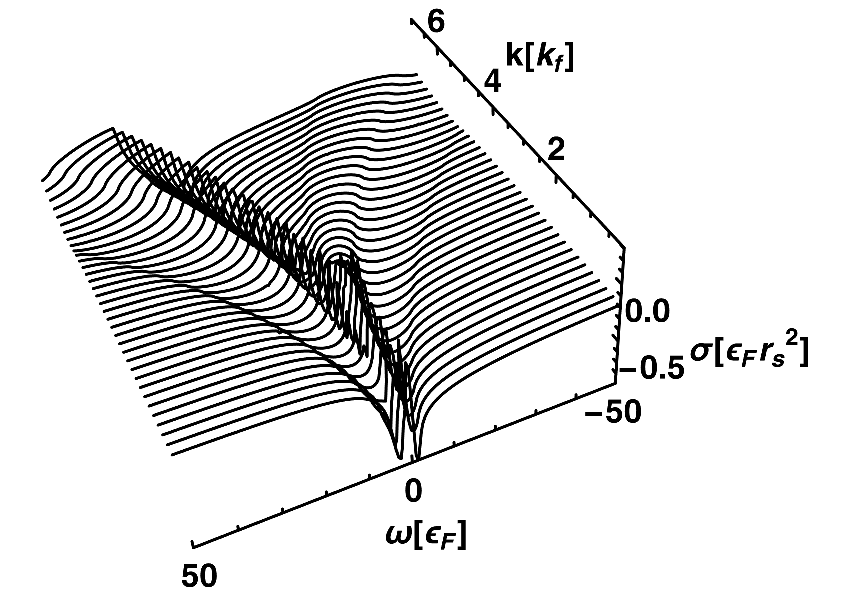}
\includegraphics[width=8cm]{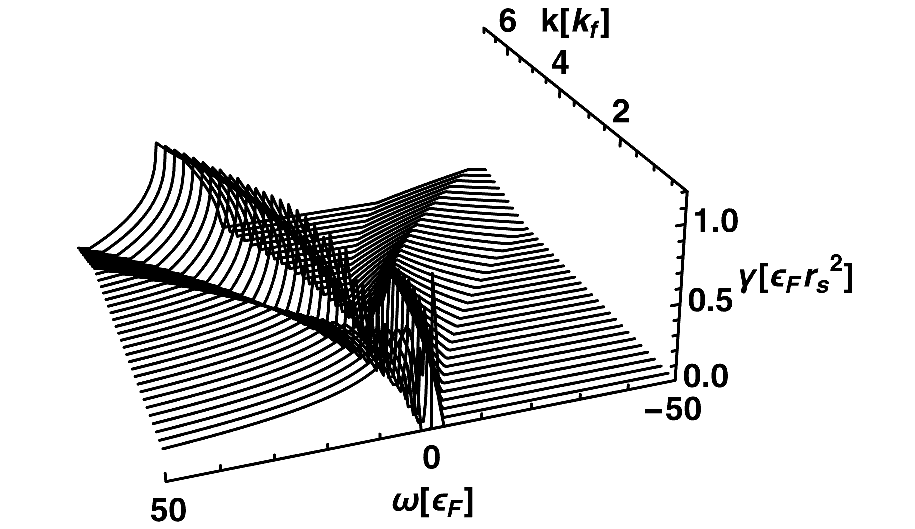}
\caption{The real part $\sigma$ (above) and imaginary part $\gamma$ (below) of the selfenergy for contact interaction.
\label{s_real}}
\end{figure}

The real and imaginary parts of the selfenergy are plotted in figure (\ref{s_real}). One sees that for $k>2k_f$ the damping vanishes in a range $\omega\approx 2$ which is accompanied with a gap as seen in figure~\ref{finitei}.
Further a splitting of two excitation lines for positive frequencies and one for negative frequencies appear. Which of them finally survives and describes a real excitation in the system is decided by the spectral function in the next Section.

\begin{figure}[h]
\includegraphics[width=8cm]{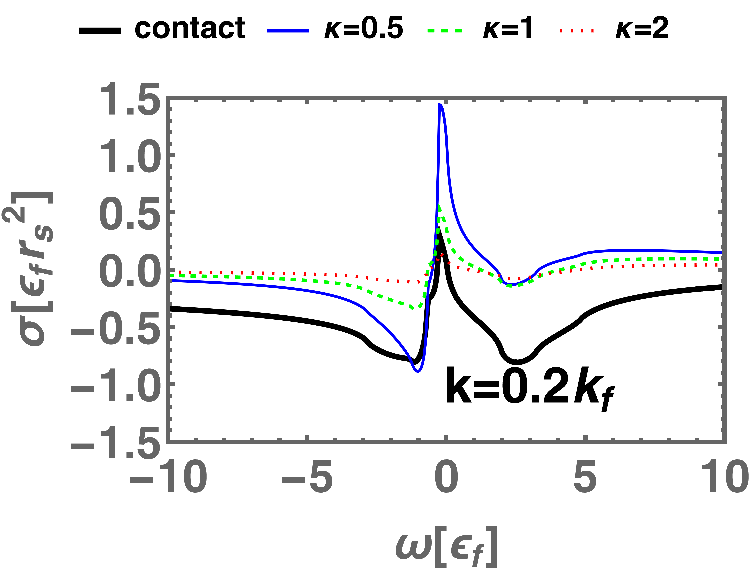}
\caption{The real part 
of the selfenergy for contact interaction and three values of finite size potential (\ref{pot}) corresponding to the imaginary part in figure \ref{finitei}.
\label{finite}
}
\end{figure}

The finite size of a potential (\ref{pot}) is compared in figure~\ref{finite} for different values of the screening parameter $\kappa$ with the contact interaction. For small $\kappa$ we approach Coulombic behaviour and see that the peak at the Fermi energy becomes enhanced. This is a similar behaviour as we have seen by the imaginary part in figure \ref{finitei}.

\section{Spectral function}

Now that we have the real and imaginary part of the selfenergy (\ref{Ga}) we can calculate the spectral function as measure for the single-particle excitation in the system
\begin{eqnarray}
a(\omega,k)={\gamma(\omega,k)\over [\omega-k^2-\sigma^F(k)-\sigma(\omega,k)]^2+{\gamma^2(\omega,k)\over 4}}.
\label{aspec}
\end{eqnarray}
The still missing part is the Hartree-Fock meanfield selfenergy as the part lower than Born in perturbation theory and it is necessary to include the meanfield in order to have all results systematically up to second order in the potential. The Hartree selfenergy proportional to the number of electrons $\sigma^H=n V_0$ is compensated by a neutralizing background. The Fock term as exchange meanfield term reads for contact interaction $v_q=1$ and finite-size potential (\ref{pot})
\begin{align}
\sigma^F(k)&=-s \int\limits_{-\infty}^\infty {d q\over 2 \pi \hbar} v_q n_{k-q}
\nonumber\\
&=-{2 s^2\over \pi^2} r_s
\left \{
\begin{array}{cc}
1& {\rm , contact}\cr &
\cr
\frac{2}{\pi ^2} \ln \left|
   \frac{k-\sqrt{(k+1)^2+\text{kap}^2}+1}{-k+\sqrt{(1-k)^2+\text
   {kap}^2}+1}\right|
& {\rm , pot.} (\ref{pot})
\end{array}
\right .,
\label{F}
\end{align} 
respectively, using the Bruckner or coupling parameter $r_s$.

\begin{figure}[h]
\includegraphics[width=7cm]{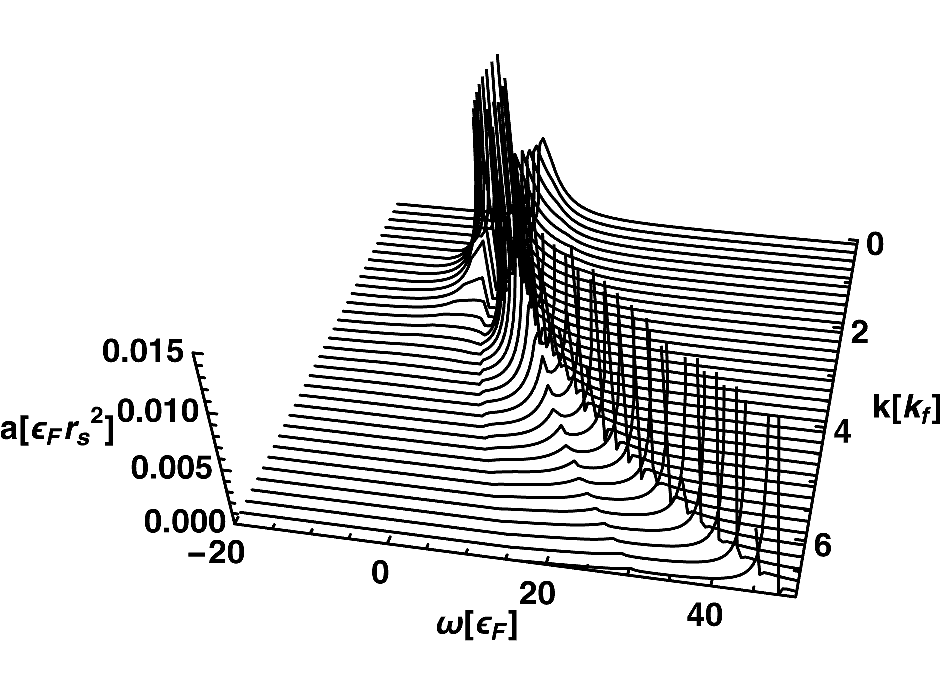}
\includegraphics[width=8cm]{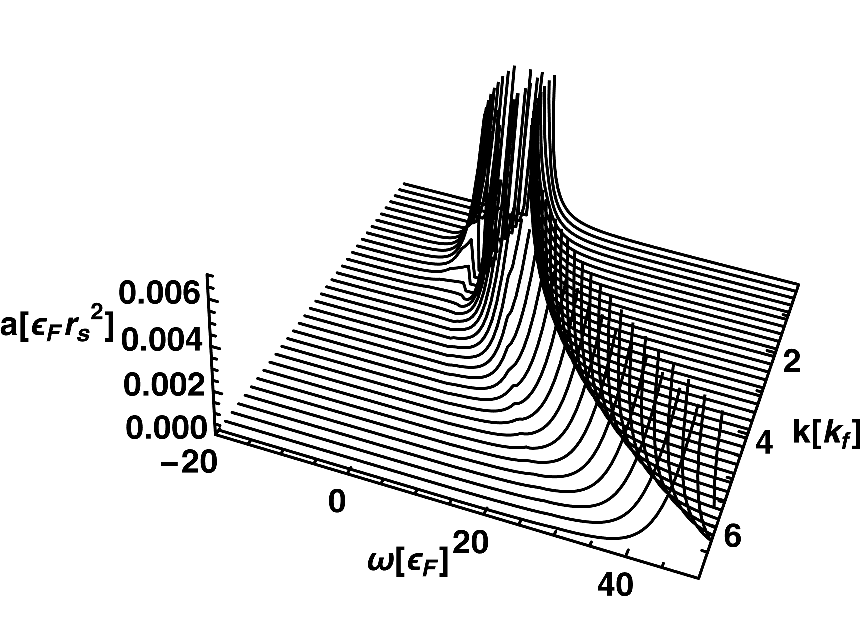}
\caption{The spectral function (\ref{aspec}) with coupling parameter $r_s=0.5$ for contact interaction (above) and for finite size (\ref{pot}) with $\kappa=0.5$ (below).\label{spect_rs05}}
\end{figure}

\begin{figure}[h]
\includegraphics[width=8cm]{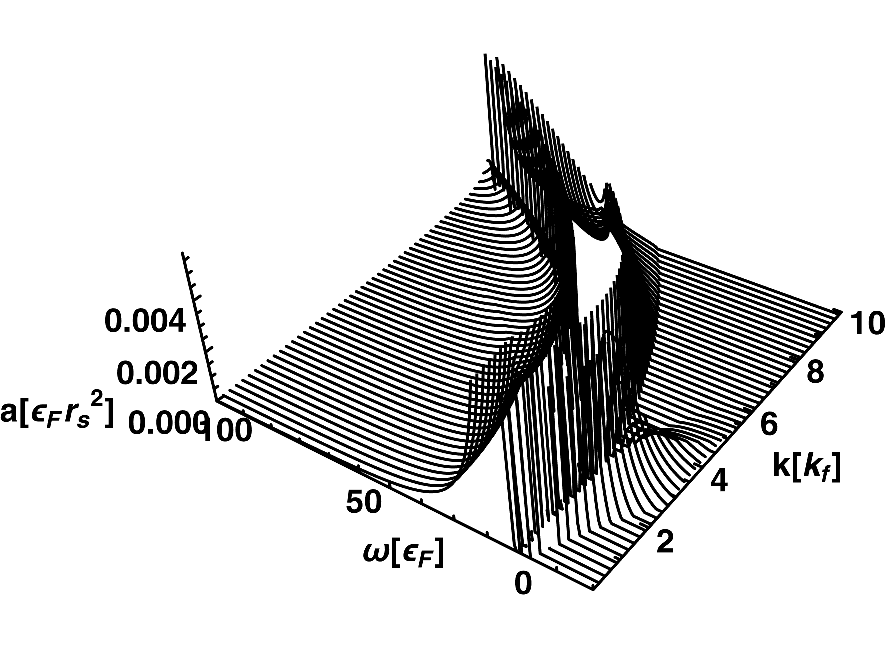}
\includegraphics[width=8cm]{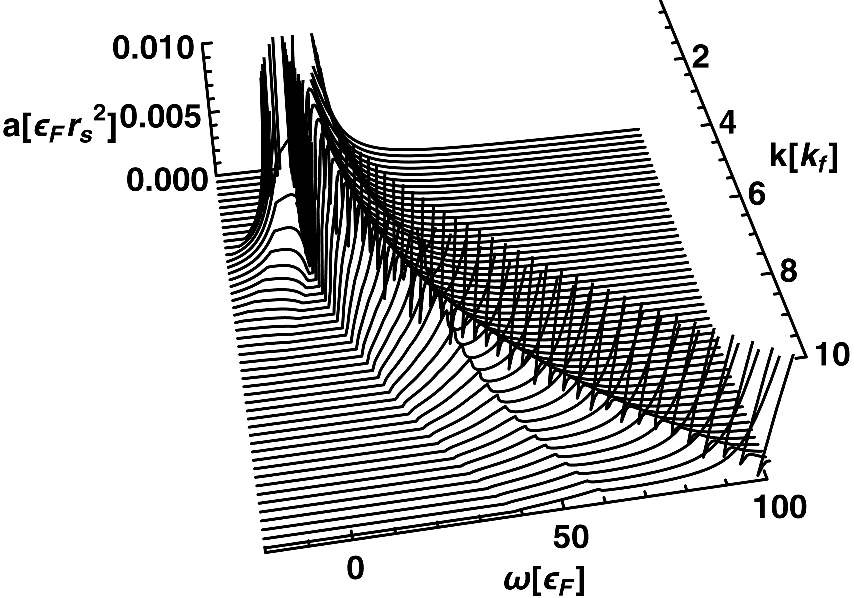}
\caption{The spectral function (\ref{aspec}) for contact interaction and a coupling parameter $r_s=2$ in different views.
\label{spect_rs2}}
\end{figure}

In figures~\ref{spect_rs05} and \ref{spect_rs2} we plot the spectral functions for a Bruckner parameter $r_s=0.5$ and $r_s=2$ respectively. One recognizes the main excitation at $\omega=k^2$ line for large momenta. For contact interaction a splitting of the quasiparticle excitation pole appears at higher momenta which is absent in the finite-size or Coulombic potential. The characteristic feature of contact potential is more clearly visible for higher coupling in figure~\ref{spect_rs2}. A gap opens with the borders $\omega\approx \pm 2 k$ (or in units $\hbar \omega \approx {p_f\over m} k$) which feature would be the exact borders for a Luttinger liquid \cite{V83,G04}. Here we do not have a Luttinger liquid but see similar features. The two peaks are related to holon and antiholon excitation's, i.e. the excitation of a particle out of Fermi sea \cite{EFGKK10}, schematically illustrated in figure~\ref{exit}. The corresponding threshold singularities have been discussed in \cite{Ess10}. The deviation from the Luttinger liquid can be seen by the boundary of the gap in figure~\ref{s_real} which should be linear $\pm k v_o$ with the charge velocity \cite{Vo93a,MeS92} of $v_0=\sqrt{v_f^2+g^2}$. A spin-polarized system would show additionally a splitting of the peak in spin and charge velocities \cite{Vo93,G04}. Further we do have a finite width of the peaks of the spectral function in contrast to the Tomonaga-Luttinger model \cite{MeS92}. The appearance of the gap is also related to a pseudogap in the density of states \cite{Ta20}.

Up to the momentum of $k=2k_f$ there appears an excitation at negative frequencies which one interprets as bound states. Momenta above $2 k_f$ correspond to nesting which means that this bound state is destroyed when nesting occurs. The occurrence of this bound state is puzzling since it appears in 3D only for higher-order approximations like the ladder summation. Since we consider the weak-coupling limit which is the high-density limit, we see probably here a precursor of bound states in the off-shell selfenergy. 

\section{Summary and Conclusions}

The Born selfenergy including exchange is expressed analytically with a remaining single integral for the imaginary part (\ref{phi}) and the real part (\ref{psi1}) respectively. This allows to calculate the selfenergy precise and fast for any interaction potential. Therefore these expressions can be applied widely. The momentum-frequency range of different parts of the selfenergy turns out to be an astonishing complex consisting of single-particle excitation and border lines of collective modes. This leads to a non-differentiable behaviour of the imaginary part of the selfenergy. The real part is worked down as well to a single integral which provides a fast scheme. Given the Born selfenergy together with the meanfield, the spectral function as measure for single-particle excitation is calculated for the illustrative examples of contact interaction and a finite-size potential. The opening of the Luttinger gap is seen with increasing momenta. Two excitation lines due to holon and antiholon excitations are observed. An excitation at negative frequencies is interpreted as a precursor of bound states which vanishes for momenta exceeding $2p_f$ instead indicating nesting.

\begin{acknowledgments}
K.N.P.\ acknowledges the grant of honorary senior scientist position by National Academy of Sciences of India (NASI) Prayagraj. 
K.M.\ acknowledges support from DFG-project
MO 621/28-1. V.A. acknowledge the support in the form of SERB-Core research Grant No. CRG/2023/001573.
\end{acknowledgments}

\appendix
\section{q-integration of $\sigma <$\label{selfs}}

Here we show an analytical way to calculate the selfenergy (\ref{Born}).
First we observe that it is only necessary to integrate half of the range $q>0$ in (\ref{Born}) since the area $q<0$ can be mapped to the $q>0$ expression if we set $k\to -k$. This can be seen in (\ref{Born}) since the $p$ integration allows to set $p\to -p$. 
The $\delta$-function in (\ref{Born}) is carried out which means to replace
\begin{eqnarray}
p=q+k-{\Omega\over 2 q},\quad \Omega=\omega-k^2
\end{eqnarray}
and providing an additional $m/|q|$ prefactor. Together with the potential, it becomes
\begin{eqnarray}
&&\delta (\omega\!+\!\epsilon_p\!-\!\epsilon_{p\!-\!q}\!-\!\epsilon_{k\!+\!q})V_q\left [s V_q-V_{p-k-q}\right ]
\nonumber\\
&&=
{\hbar^4\over m a_B^2}\delta\left (p-q-k+{\Omega\over 2 q}\right ){v_q\over |q|} \left (s v_q-v_{\Omega\over 2 q}\right ).
\end{eqnarray} 

\begin{figure}[h]
\includegraphics[width=8cm]{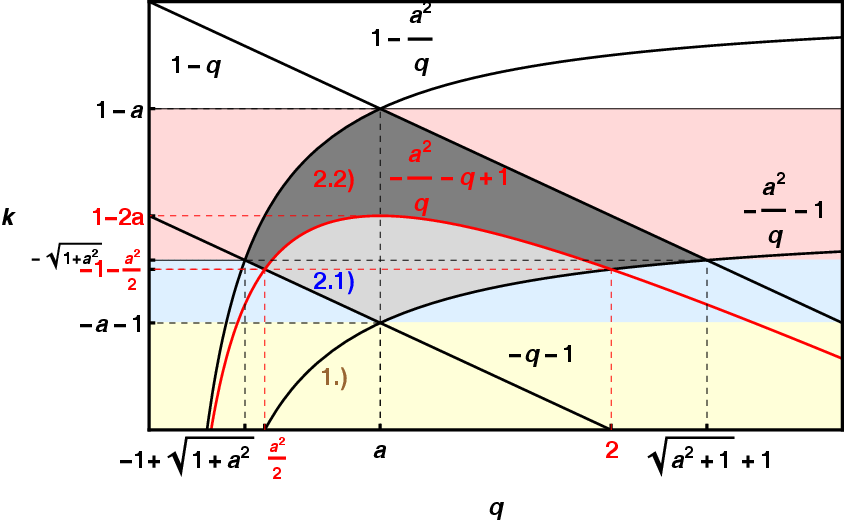}
\caption{The condition (\ref{cond1}) for the allowed region (light gray) bounded by $-1-q,1-q,1-a^2/q, -1-a^2/q$ and additionally to be larger than $1-a-a^2/q$ (gray). Depending on the maxima of the latter function (red) at $(a,1-2a)$ there are three cases, 1.2, 2.1,and 2.2. 
\label{ss_123}}
\end{figure}

From occupation factors in (\ref{Born}) we get the conditions $\Omega=\omega-k^2$
\begin{eqnarray}
&n_{p-q}:& -1+{\Omega \over 2 q}<k<1+{\Omega \over 2 q}
\nonumber\\
&n_{k+q}:&-1-q<k<1-q 
\nonumber\\
&1-n_{p}:&k<-1-q+{\Omega\over 2 q}\, {\rm or}\, 1-q+{\omega\over 2 q}<k
\end{eqnarray}
which allows two possibilities for the range of $k$ 
\begin{align}
&{\rm max}\!\left (\!-1\!-\!q,\!-1\!+\!{\Omega\over 2 q},\!1\!+\!{\Omega\over 2 q}\!-\!q\!\right )\!<\!k\!<\!{\rm min}\!\left (\!1\!-\!q,\!1\!+\!{\Omega\over 2 q}\!\right )
\nonumber\\
&{\rm or}
\nonumber\\
&{\rm max}\!\left (\!\!-1\!-\!q,\!-1\!+\!{\Omega\over 2 q}\!\right )\!\!<\!k\!<\!{\rm min}\!\left (\!1\!-\!q,\!1\!+\!{\Omega\over 2 q},\!-1\!+\!{\Omega\over 2 q}\!-\!q\!\right ).
\end{align}
Since $q>0$ we see that the second line is not possible to complete since it would require $-1\!+\!{\Omega\over 2 q}<-1+{\Omega\over 2 q}-q$. 
From the first line we see that
$\Omega<0$ since otherwise $1\!+\!{\Omega\over 2 q}\!-\!q<1-q$ would be impossible. Therefore setting 
\begin{eqnarray}
a=\sqrt{-\Omega/2}
\label{aa}
\end{eqnarray}
we have to discuss
\begin{align}
{\rm max}\!\left (\!-1\!-\!q,\!-1\!-\!{a^2\over q},\!1\!-\!{a^2\over q}\!-\!q\!\right )\!<\!k\!<\!{\rm min}\!\left (\!1\!-\!q,\!1\!-\!{a^2\over q}\!\right )
\label{cond1}
\end{align}
which is plotted in figure~\ref{ss_123}. 

\begin{figure}[h]
\includegraphics[width=8cm]{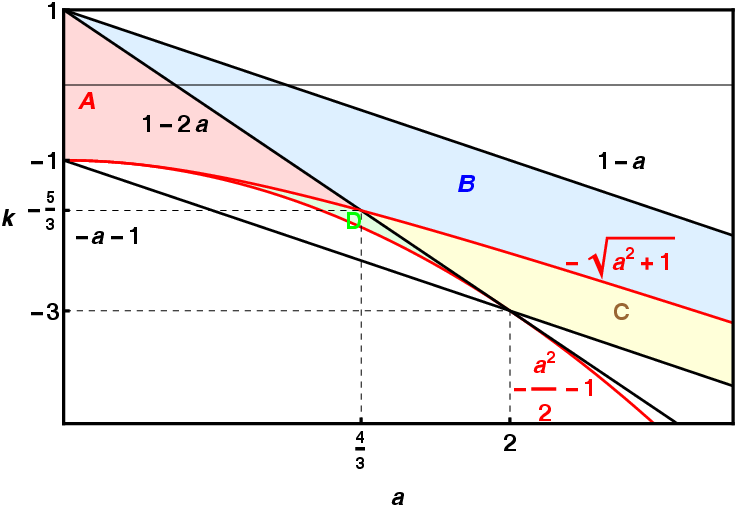}
\caption{The condition (\ref{cond1}) of figure \ref{ss_123} rearranged in $a-k$ plot yielding 4 different regions.
\label{ss_ak}}
\end{figure}

The allowed region is in a four-polygon and additionally above the curve $1-a^2/q-q$. Due to its maxima at $(a,1-2a)$ we have 3 cases:\\ 
\begin{itemize}
\item[(1.)] $1-2 a< -1-a$ which means $a>2$ and we have
\begin{eqnarray}
&&-1-a<k<-\sqrt{1+a^2}: -1-k<q<-{a^2\over 1+k}\nonumber\\
&& -\sqrt{1+a^2}<k<1-a: {a^2\over 1-k}<q<1-k
\end{eqnarray}
\item[(2.1)] $-1-a<1-2 a$ which means $\frac 4 3 < a <2$ and
\begin{eqnarray}
&&-1-{a^2\over 2}<k<1-2 a:\nonumber\\
&&\qquad\qquad  -1-k<q<q_2\quad{\rm or}\quad q_1<q<-{a^2\over 1+k}\nonumber\\
&&1-2 a<k<-\sqrt{1+a^2}: -1-k<q<-{a^2\over 1+k}\nonumber\\
&& -\sqrt{1+a^2}<k<1-a: {a^2\over 1-k}<q<1-k
\end{eqnarray}
\item[(2.2)] $-\sqrt{1+a^2}<1-2a < 1-a$ which yields $0<a<\frac 4 3$ and
\begin{eqnarray}
&&-1-{a^2\over 2}<k<-\sqrt{1+a^2}:\nonumber\\
&&\qquad\qquad  -1-k<q<q_2 \quad{\rm or}\quad q_1<q<-{a^2\over 1+k}\nonumber\\
&&-\sqrt{1+a^2}<k<1-2 a:\nonumber\\
&&\qquad\qquad  {a^2\over 1-k}<q<q_2 \quad{\rm or}\quad q_1<q<1-k\nonumber\\
&& -\sqrt{1+a^2}<k<1-a: {a^2\over 1-k}<q<1-k
\label{12}
\end{eqnarray}
\end{itemize}
with the two crossing points of the $1-q-a^2/q$ curve with the horizontal $k$-line
\begin{eqnarray}
q_{1/2}={1-k\over 2}\pm \sqrt{{(1-k)^2\over 4}-a^2}.
\label{q12}
\end{eqnarray}

Eq. (\ref{12}) provides the integration limits for $q$ of
\begin{align}
\Phi(q)&= {s^4 r_s^2\over \pi^3}\int\limits^qd \bar q{v_{\bar q}\over |\bar q|}\left (s v_q-v_{a^2\over q}\right )
\nonumber\\
&= {s^4 r_s^2\over \pi^3}\left \{
\begin{array}{cc}
\ln q&{\rm contact}
\cr
\frac{\ln \left(\frac{q^2}{\kappa ^2+q^2}\right)}{\kappa ^2}+\frac{i F\left(i {\rm arsinh}\left(\frac{q}{\kappa }\right)|\frac{\kappa
   ^4}{a^4}\right)}{a^2}&{\rm pot.} (\ref{pot}) 
\end{array}
\right ..
\label{phia}
\end{align}
with the elliptic integral of first kind 
\begin{eqnarray}
F(a|m)=\int\limits_0^a{d\theta\over \sqrt{1-m\sin^2\theta}}.
\end{eqnarray}

Plotting the cases 1., 2.1, and 2.2. and regrouping with respect to $k$ one sees in figure~\ref{ss_ak} that 4 areas $A-D$ appear with 3 combinations of 
\begin{align}
&{\rm area}\, A,B:\, &\Phi(1-k)-\Phi\left ({a^2\over 1-k}\right )
\nonumber\\
&{\rm area}\, A,D:\, &\Phi(q_2)-\Phi(q_1)
\nonumber\\
&{\rm area}\, C,D:\, &\Phi\left ({a^2\over -1-k}\right)-\Phi(-1-k)
\end{align}
Remembering that in order to include the $q<0$ part we have to add all expressions for $k\to -k$ which provides finally the cases (\ref{forma}). 

\section{$q$-integration for $\sigma^>$\label{selfg}}

The particle contribution $\sigma^>$ (\ref{Bornl}) to the selfenergy is now calculated as in the appendix before.
The occupation factors in (\ref{Bornl}) lead after $\delta$-integration for $p$ to the conditions
\begin{eqnarray}
&1-n_{p-q}:& k<-1+{\Omega \over 2 q}\quad {\rm or}\quad 1+{\Omega \over 2 q}<k
\nonumber\\
&n_{k+q}:&k<-1-q\quad {\rm or}\quad 1-q<k 
\nonumber\\
&n_{p}:&-1-q+{\Omega\over 2 q}<k<1-q+{\omega\over 2 q}<k
\end{eqnarray}
which together allows four possibilities for the range of $k$ 
\begin{align}
&{\rm max}\!\left (\!-1\!+\!{\Omega\over 2 q}\!-\!q\!\right )\!<\!k\!<\!{\rm min}\!\left (-\!1\!-\!q,\!1\!+\!{\Omega\over 2 q}\!-\!q,-1\!+\!{\Omega\over 2 q}\!\right )
\label{cond2}\\
&{\rm or}
\nonumber\\
&{\rm max}\!\left (\!-1\!+\!{\Omega\over 2 q}\!-\!q,1\!+\!{\Omega\over 2 q}\!\right )\!<\!k\!<\!{\rm min}\!\left (-\!1\!-\!q,\!1\!+\!{\Omega\over 2 q}\!-\!q\!\right )
\nonumber\\
&{\rm or}
\nonumber\\
&{\rm max}\!\left (\!-1\!+\!{\Omega\over 2 q}\!-\!q,1-q\!\right )\!<\!k\!<\!{\rm min}\!\left (\!-1\!+\!{\Omega\over 2 q},1\!+\!{\Omega\over 2 q}\!-\!q\!\right )
\label{cond3}\\
&{\rm or}
\nonumber\\
&{\rm max}\!\left (\!-1\!+\!{\Omega\over 2 q}\!-\!q,1-q,1\!+\!{\Omega\over 2 q},\!\right )\!<\!k\!<\!{\rm min}\!\left (\!1\!+\!{\Omega\over 2 q}\!-\!q\!\right )
\nonumber
\end{align}
where the second and fourth line is not possible to complete due to the last expressions on left and right side $1\!+\!{\Omega\over 2 q}\!<\!k\!<\!1\!+\!{\Omega\over 2 q}\!-\!q$.
The first line is only possible to complete for $\Omega<0$ since $-1\!+\!{\Omega\over 2 q}\!-\!q\!<\!k\!<\!-\!1\!-\!q$ and the third line only for $\Omega>0$ due to $1\!-\!q\!<\!k\!<\!1\!+\!{\Omega\over 2 q}\!-\!q\!$.
Therefore we have two cases: 

\subsubsection{$\Omega=\omega-k^2<0$}

\begin{figure}[h]
\includegraphics[width=8cm]{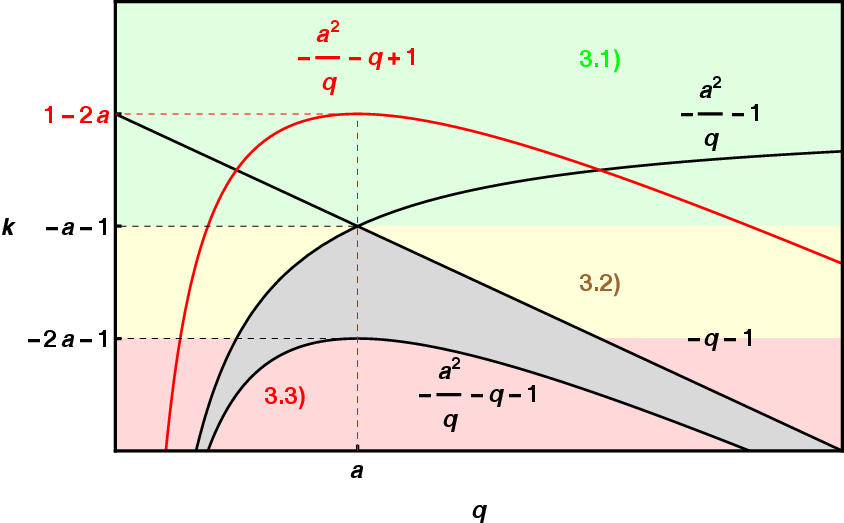}
\caption{The condition (\ref{cond2}) for the allowed region (light gray) bounded by $-1-q,-1-a^2/q, -1-a^2/q-q$ and additionally to be smaller than $1-q-a^2/q$ (red line). Depending on the maxima of the latter function (red) at $(a,1-2a)$ there are three cases (3.1)-(3.3). 
\label{sl_123}}
\end{figure}

Discussing first the case (\ref{cond2}) and setting 
\begin{eqnarray}
a=\sqrt{-\Omega/2}
\label{aa1}
\end{eqnarray}
the area is plotted in figure~\ref{sl_123}. 

\begin{figure}[h]
\includegraphics[width=8cm]{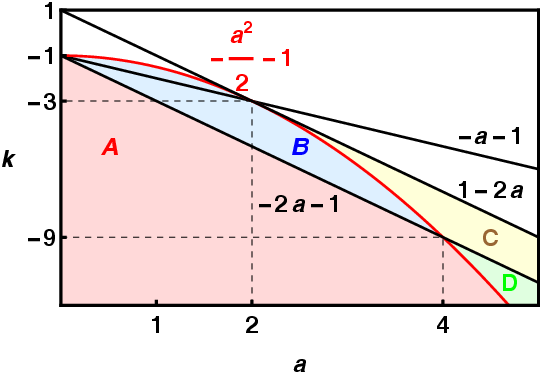}
\caption{The condition (\ref{cond2}) of figure \ref{sl_123} rearranged in $a-k$ plot yielding 4 different regions.
\label{sl_ak}}
\end{figure}

The allowed region is additionally below the curve $-1-a^2/q-q$. Due to its maxima at $(a,1-2a)$ we have 3 cases:\\ 
\begin{itemize}
\item[(3.1)] 
$1-2 a< -1-a$ which means $a>2$ and we have
\begin{eqnarray}
&& -\infty<k<-1-2 a:\nonumber\\&&\qquad -{a^2\over 1+k}<q<q_4\quad{\rm or}\quad q_3<q<-1-k
\nonumber\\
&& -1\!-\!2 a<k<-1\!-\!a: -{a^2\over 1\!+\!k}<q<-1\!-\!k
\end{eqnarray}
\item[(3.2)] $-1-2a<-1-a$ and $-1-a^2/2>-1-2 a$ which means $2 < a <4$ and
\begin{eqnarray}
&&-\infty<k<-1-2 a:\nonumber\\&&\qquad
-{a^2\over 1+k}<q<q_4\quad{\rm or}\quad q_3<q<-1-k\nonumber\\
&&-1-2 a<k<-1-{a^2\over 2}: -{a^2\over 1+k}<q<-1-k\nonumber\\
&& -1-{a^2\over 2}<k<1-2a: q_2<q<q_1
\end{eqnarray}
\item[(3.3)] $-1-{a^2\over 2} < -1-2 a$ which yields $4<a$ and
\begin{eqnarray}
&&-1-{a^2\over 2}<k<-1-2 a:\nonumber\\
&&\qquad  q_2<q<q_4\quad{\rm or}\quad q_3<q<q_1\nonumber\\
&&-1-2 a<k<1-2 a: q_2<q<q_1\nonumber\\
&& -\infty<k<-1-{a^2\over 2}: \nonumber\\&&\qquad-{a^2\over 1-k}<q<q_4\quad{\rm or}\quad q_3<q< -1-k
\label{caseqk}
\end{eqnarray}
\end{itemize}
with the two crossing points of the $\pm 1-q-a^2/q$ curves with the horizontal $k$-line of (\ref{q12}) 
\begin{eqnarray}
q_{3/4}={-1-k\over 2}\pm \sqrt{{(1+k)^2\over 4}-a^2}.
\label{q34}
\end{eqnarray}

In figure~\ref{sl_ak} we plot these ranges and obtain 4 areas $A-D$ with 3 combinations of 
\begin{align}
&{\rm area}\, A,B:\, &\Phi(-1-k)-\Phi\left ({a^2\over -1-k}\right )
\nonumber\\
&{\rm area}\, A,D:\,&\Phi(q_4)-\Phi(q_3)
\nonumber\\
&{\rm area}\, C,D:\, &\Phi\left (q_1\right)-\Phi(q_2)
\end{align}

\begin{figure}[h]
\includegraphics[width=8cm]{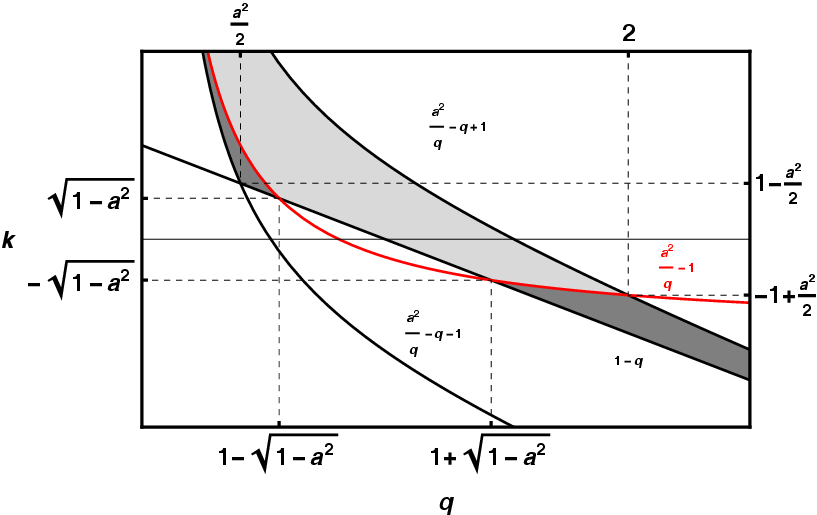}
\includegraphics[width=8cm]{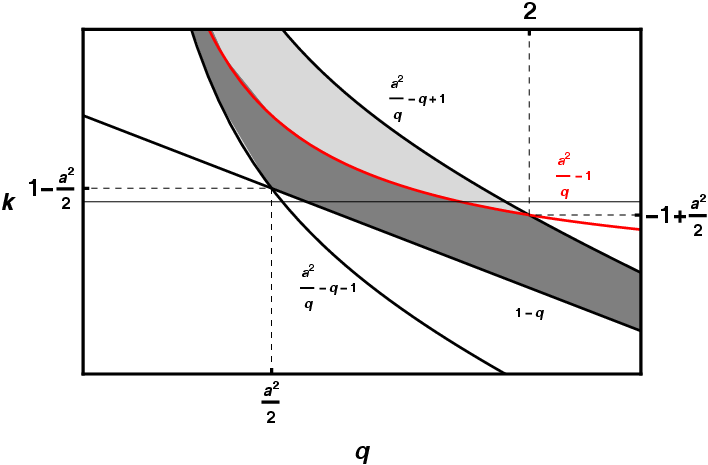}
\includegraphics[width=8cm]{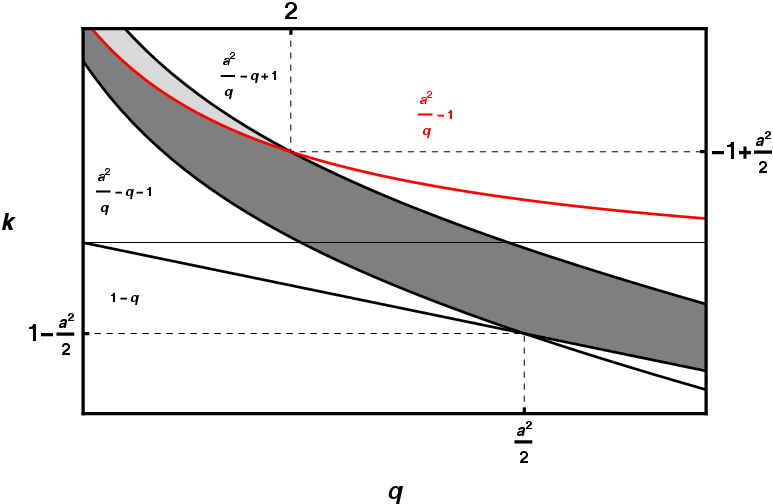}
\caption{The condition (\ref{cond3}) for the allowed region (light gray) bounded by $-1-q,-1-a^2/q, -1-a^2/q-q$ and additionally to be smaller than $a^2/q-1$ (gray). Depending on the latter function (red) there are three cases 4.1{)}-4.3{)} from top to bottom. 
\label{sl_41}}
\end{figure}

\begin{figure}[h]
\includegraphics[width=8cm]{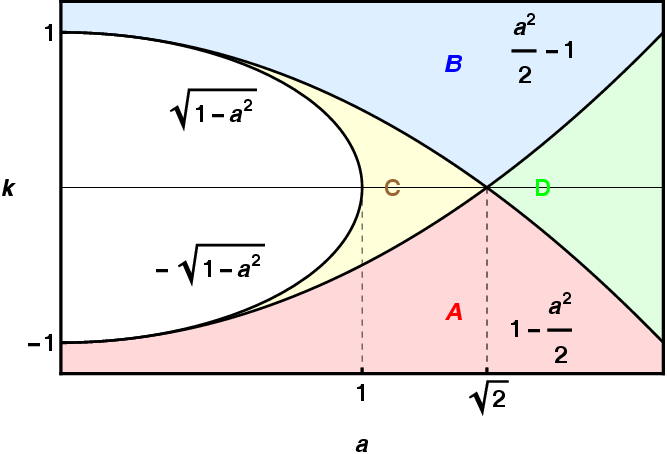}
\caption{The condition (\ref{cond2}) of figure \ref{sl_41} rearranged in $a-k$ plot yielding 4 different regions.
\label{sl4_ak}}
\end{figure}

Again we have to add all expressions for $k\to -k$ to get finally the cases (\ref{formb})

\subsubsection{$\Omega=\omega-k^2>0$}

The condition (\ref{cond3}) and setting $a=\sqrt{\Omega/2}$ can be seen in figure~\ref{sl_41}. 
The allowed region is additionally below the curve $-1-a^2/q-q$ and we have 3 cases:\\ 
4.1{)} ${a^2\over 2}<1-\sqrt{1-a^2}<1+\sqrt{1-a^2}<2$ which means $a<1$ and we have
\begin{eqnarray}
&& -\infty<k<-1+{a^2\over 2}: 1-k<q<q_7
\nonumber\\
&& -1\!+\!{a^2\over 2}<k<-\sqrt{1-a^2}: 1\!-\!k<q<{a^2\over 1\!+\!k}
\nonumber\\
&& \sqrt{1-a^2}<k<1\!-\!{a^2\over 2}: 1\!-\!k<q<{a^2\over 1\!+\!k}
\nonumber\\
&& 1\!-\!{a^2\over 2}<k<\infty: q_8<q<{a^2\over 1\!+\!k}
\end{eqnarray}
4.2{)} $1-a^2/2>-1+a^2/2>-\sqrt{1-a^2}$ which means $1 < a <\sqrt{2}$ and
\begin{eqnarray}
&& -\infty<k<-1+{a^2\over 2}: 1-k<q<q_7
\nonumber\\
&& -1\!+\!{a^2\over 2}<k<1\!-\!{a^2\over 2}: 1\!-\!k<q<{a^2\over 1\!+\!k}
\nonumber\\
&& 1\!-\!{a^2\over 2}<k<\infty: q_8<q<{a^2\over 1\!+\!k}
\end{eqnarray}
4.3{)} $1-a^2/2<-1+a^2/2$ which means $\sqrt{2}<a$ and
\begin{eqnarray}
&& -\infty<k<1-{a^2\over 2}: 1-k<q<q_7
\nonumber\\
&& 1\!-\!{a^2\over 2}<k<-1\!+\!{a^2\over 2}: q_8<k<q<q_7
\nonumber\\
&& -1\!+\!{a^2\over 2}<k<\infty: q_8<q<{a^2\over 1\!+\!k}
\end{eqnarray}
with the two crossing points of the $\pm 1-q+a^2/q$ curve with horizontal line 
\begin{eqnarray}
q_{7/8}={\pm 1-k\over 2}+ \sqrt{{(1\mp k)^2\over 4}+a^2}.
\label{q78}
\end{eqnarray}

In figure~\ref{sl4_ak} we plot these ranges and obtain 4 areas $A-D$ with the combinations 
\begin{align}
&{\rm area}\, A\,: &\Phi\left (q_7\right)-\Phi(1-k)
\nonumber\\
&{\rm area}\, B\,:&\Phi\left ({a^2\over 1+k}\right )-\Phi(q_8)
\nonumber\\
&{\rm area}\, C\,:&\Phi\left ({a^2\over 1+k}\right )-\Phi(1-k)
\nonumber\\
&{\rm area}\, D\,:&\Phi\left (q_7\right)-\Phi(q_8).
\end{align}

Again we have to add all expressions for $k\to -k$ to get the cases (\ref{formc}).

\section{Real part of selfenergy\label{reals}}

We calculate the Hilbert transform (\ref{Hilbert}) by interchanging integration orders 
\begin{eqnarray} 
\sigma(q,\omega)&=&\int\limits_{-\infty}^\infty {d\bar \omega \over 2 \pi}{\sigma^<(q,\bar \omega)\over \omega-\bar \omega}
\nonumber\\
&=&\int\limits_0^\infty{d x\over 2 \pi} {\sigma^<(q,x)\over {\cal O}+x}+
\int\limits_{-\infty}^\infty{d x\over 2 \pi} {\sigma^>(q,x)\over {\cal O}-x}
\nonumber\\
&=&{s^4 r_s^2\over 2 \pi^4}\int\limits_u^o{d q}{v_q\over q}\int\limits_{x_u}^{x_o}{dx \over {\cal O}+x}\left (s v_q-v_{x\over q}\right )
\label{Hilberts}
\end{eqnarray}
where the integration limit of $q$ according to (\ref{forma}) are interchanged with $a^2=x=(k^2-\omega)/2$ integration leading to the limits summarized in table~\ref{limit}.
We have abbreviated 
\begin{eqnarray}
\w=(\omega-k^2)/2\gtrless 0
\end{eqnarray} 
and used a transformation of $x\to -x$ in the part for $\sigma^>$.

As example, the case
\begin{align}
&0<k<1&\Phi(1-k)-\Phi\left ({a^2\over 1-k}\right )\,{\rm for}\,&0\!<\! a^2\!<\! (1\!-\!k)^2
\end{align}
leads to
\begin{align}
&\int\limits_0^{(1-k)^2}{d x\over \w+x} \int\limits_{x\over 1-k}^{1-k} {dq\over q}v_q(s v_q-v_{x\over q}) 
\nonumber\\
&=\int\limits_{0}^{1-k} {dq\over q}v_q \int\limits_0^{(1-k) q}{d x\over \w+x}(s v_q-v_{x\over q}) 
\label{a}
\end{align}
as represented in first line of table~\ref{limit}.
Working out all cases of (\ref{forma}) is tedious but straight by just painting all the corresponding curves. Summing up the contributions according to the range of $k$
we obtain (\ref{real}) which shows that the ranges $1<k<3$ and $k>3$ are identically as it should since we have only $k=1(k_f)$ as exceptional point.

\begin{table}[h]
\footnotesize
\caption{Integration limits of \ref{Hilberts} where we use the abbreviations $[ij]=-q^2+i kq+jq$ and $(ij)=i kq+jq$.}\label{limit}
\begin{tabular}{|c|c|c|c|c|}
\hline
$\sigma^<$ &$ u $&$ o $&$x_u$ & $x_o$\\
\hline
$0<k<1$& &&&\\
\hline
&$ 0$&$1-k$&$0$&$(-+)$\\
\hline
&$1-k$&$0$&$0$&$[-+]$\\
\hline
&$ 0$&$1+k$&$0$&$(++)$\\
\hline
&$1+k$&$0$&$0$&$[++]$\\
\hline
$1<k<3$&&&&\\
\hline
&$2$&$k+1$&$(+-)$&$k^2-1$\\
\hline
&$k-1$&$2$&$[++]$&$(++)$\\
\hline
&$k-1$&$k+1$&$k^2-1$&$(++)$\\
\hline
$3<k$&&&&\\
\hline
&$k-1$&$k+1$&$(--)$&$(+-)$\\
\hline
\end{tabular}

\vspace*{0.5cm}

\begin{tabular}{|c|c|c|c|c|}
\hline  
$\sigma^>$ &$ u $&$ o $& $x_u$ & $x_o$\\
\hline
$0<k<1$& &&&\\
\hline
&$ 1+k$&$2$&$[--]$&$(--)$\\
\hline
&$1-k$&$1+k$&$(+-)$&$(--)$\\
\hline
&$ 2$&$\infty$&$[--]$&$[++]$\\
\hline
&$ 2$&$\infty$&$[-+]$&$[+-]$\\
\hline
$1<k<3$&&&&\\
\hline
&$0$&$2$&$0$&$(--)$\\
\hline
&$1+k$&$2$&$[-+]$&$0$\\
\hline
&$0$&$k+1$&$[--]$&$0$\\
\hline
&$k+1$&$\infty$&$[--]$&$[++]$\\
\hline
&$k+1$&$\infty$&$[+-]$&$[-+]$\\
\hline
&$0$&$k-1$&$[+-]$&$(+-)$\\
\hline
$3<k$&&&&\\
\hline
&$0$&$2$&$0$&$(--)$\\
\hline
&$1+k$&$2$&$[-+]$&$0$\\
\hline
&$0$&$k+1$&$[--]$&$0$\\
\hline
&$k+1$&$\infty$&$[--]$&$[++]$\\
\hline
&$k+1$&$\infty$&$[+-]$&$[-+]$\\
\hline
&$0$&$k-1$&$[+-]$&$0$\\
\hline
&$0$&$2$&$0$&$(+-)$\\
\hline
&$2$&$k-1$&$2(k-1)$&$[++]$\\
\hline
\end{tabular}
\end{table}

\newpage


\end{document}